\documentstyle[12pt]{article}
\begin{document}
\begin{center}
ALL-TO-ALL CHAOS SYNCHRONIZATION IN A NETWORK OF NETWORKS:ONE OF THE SIMPLEST CASES\\
E.M.Shahverdiev$^{1}$, P.A.Bayramov, R.A.Nuriev, L.H.Hashimova and M.V. Qocayeva\\
Institute of Physics, H.Javid,33,Baku,AZ1143,Azerbaijan\\
1. Corresponding author's e-mail:shahverdiev@physics.ab.az\\
~\\
ABSTRACT\\
\end{center}
~\\
We report on all- to- all chaos synchronization in a network of networks based on the Ikeda model. We consider one of the simplest cases. We find the existence and stability conditions for such a synchronization regime. Numerical simulations validate the analytical findings. The results can be of certain importance in achieving high level output for the coupled systems and information processing.\\ 
~\\
PACS number(s):05.45.Xt, 05.45.Vx, 42.55.Px, 42.65.Sf\\
Key words: Network of networks, Ikeda model, time-delay systems, all-to-all chaos synchronization; existence and stability conditions.\\
\begin{center}
I.INTRODUCTION
\end{center}
\indent Networks or a network of networks is a widespread concept in a world-wide-web, population dynamics, neuroscience, power grids, communication, social and computer systems, etc. Research of such interacting systems is a very hot topic in nonlinear dynamics, see e.g. [1] and references there-in. Synchronization in such systems is of certain importance in governing and performance improving point of view, e.g. enhancing emission power from such systems [1-9]. Additionally, from the fundamental point of view synchronization of coupled (chaotic) systems eliminates some degrees of freedom of the coupled system and so produces a significant reduction of complexity, thus allowing for significant simplification of computational and theoretical analysis of the system.\\
\indent As synchronization in a wider sense is associated with communication, a study of existence and stability conditions for synchronization is of paramount importance in networks. Synchronization is important in chaos based communication system to decode the transmitted message [1-9]: At the transmitter part of the communication system a message is masked with chaos, then chaos masked message is transmitted to the receiver system. At the receiver part of the communication system due to the chaos  synchronization between the transmitter and the receiver systems chaos is regenerated. Finally, deducting the receiver input and the receiver output one can decode the transmitted message, Figure 1.  \\
\indent In this paper we study one of the simplest cases of the network of networks based on the Ikeda system-paradigmatic model of chaotic dynamics. We derive analytically the existence and sufficient stability conditions for complete synchronization between all the constituents of the network. We support our analytical findings with the numerical simulations.\\
\indent The organization of the rest of this paper is as follows. In Sec. II we introduce our model. In Sec. III we present the results of analytical study. Section IV is dedicated to  the numerical simulations of all-to-all chaos synchronization between the Ikeda models. We summarize our results in Sec. IV.\\
\begin{center}
II. SYSTEM MODEL
\end{center}
\indent Consider  all-to-all synchronization between the  chaotic Ikeda systems with the following coupling topology ( see, Figure 2):x-Ikeda system governs both networks (($y,z$ and $u,w$)) which consists of only two unidirectionally coupled Ikeda systems. For simplicity consider the case when all the Ikeda systems are identical.
\begin{equation}
\frac{dx}{dt}=-\alpha x + m_{1} \sin x_{\tau}
\end{equation}

\begin{equation}
\frac{dy}{dt}=-\alpha y + m_{2} \sin y_{\tau} + m_{6} \sin x_{\tau_{1}}
\end{equation}

\begin{equation}
\frac{dz}{dt}=-\alpha z + m_{3} \sin z_{\tau} + m_{8} \sin y_{\tau_{1}}
\end{equation}

\begin{equation}
\frac{du}{dt}=-\alpha u + m_{4} \sin u_{\tau} + m_{7} \sin x_{\tau_{1}}
\end{equation}

\begin{equation}
\frac{dw}{dt}=-\alpha w + m_{5} \sin w_{\tau} + m_{9} \sin u_{\tau_{1}}
\end{equation}

\indent Here $x_{\tau} \equiv x(t-\tau)$.  The same is valid for the other dynamical variables $y,z,u,w$.
Initally the Ikeda model was introduced to describe the dynamics of an optical bistable resonator, playing an important role in electronics and physiological studies and is well-known for delay-induced chaotic behavior, see e.g.[10,11] and references there-in. Later it was established that the Ikeda model or its modifications can be used to describe the dynamics of an opto-electronical, an acousto-optical  systems and even the dynamics of the wavelength of the Distributed Bragg Reflector (DBR) Laser [11].Furthermore, this investigation is of considerable practical importance, as the equations of the class B lasers with feedback (typical representatives of class B are solid-state,
semiconductor, and low pressure $CO_{2}$ lasers [12]) can be reduced to an equation of the
Ikeda type [13].\\ 
Physically $x$ is the phase lag of the electric field across the resonator (it should be noted that in the opto-electronical and acousto-optical systems $x$ is proportinal to the voltage fed to a modulator [11]); $\alpha$ is the relaxation coefficient for the driving $x$ and driven $y,z,u,w $ dynamical variables; $\tau$ is the feedback loop time delay; 
$\tau_{1}$ is the coupling time delay between $x$ and $y$,$y$ and $z$, $x$ and $u$,$u$ and $w$; Below we will consider the case $\tau=\tau_{1}$;$m_{1},m_{2},m_{3},m_{4},m_{5}$ are the feedback strengths for the Ikeda systems $x,y,z,u,w$ respectively;
$m_{6},m_{8},m_{7},m_{9}$ are the coupling strengths between the systems $x$ and $y$,$y$ and $z$, $x$ and $u$, $u$ and $w$, respectively. 
It is noted that system $x$ is directly connected to system $y$ and connection to system $z$ occurs via system $y$. Analogously, system $x$ is directly connected to system $u$ and connection to system $w$ occurs via system $u$. It should also be emphasized that there is no direct connection between the networks ($y,z$) and ($u, w$). \\
\indent As mentioned above we will consider the all-to-all synchronization for the coupling topology presented in Fig.2.
First we consider the complete synchronization case between the variables $x$ and $y$. It is straightforward to establish that the synchronization error $\Delta_{x,y}=x-y$  under the condition 
\begin{equation}
m_{2}=m_{1}-m_{6}
\end{equation}
obeys the dynamics
\begin{equation}
\frac{d\Delta_{x,y}}{dt}= -\alpha\Delta_{x,y}+ m_{2}\Delta_{x,y} \cos x_{\tau}
\end{equation}
Obviously $\Delta_{x,y}= 0$ is a solution of system (6).\\
The sufficient stability condition of the synchronization regime 
\begin{equation}
x=y 
\end{equation}
can be found by applying the Lyapunov-Krasovskii functional approach [14,15]:
 \begin{equation}
\alpha > \vert m_{2} \vert
\end{equation}
\indent By applying this procedure to synchronization between the dynamical variables $y$ and $z$, $x$ and $z$,$x$ and $u$,
$u$ and $w$, $x$ and $w$,$y$ and $u$, $z$ and $u$,$y$ and $w$, $z$ and $w$ we establish that for the configuration in Fig.1 all-to-all complete synchronization 
\begin{equation}
x=y=z=u=w 
\end{equation}
occurs under the following conditions:
\begin{equation}
m_{1}=2 m_{2},
m_{2}=m_{3}=m_{4}=m_{5}=m_{6}=m_{7}=m_{8}=m_{9}
\end{equation}
Thus we have derived both existence (11) and stability (9) conditions for all-to-all complete synchronization (10).
In the next Section we present the results of the numerical simulations of this synchronization regime.\\
\begin{center}
III. NUMERICAL SIMULATIONS
\end{center}
\indent In this Section we numerically demonstrate how the analytical findings of the previous Section are validated. Synchronization quality is characterized by the cross-correlation coefficient $C$ [16] between the dynamical variables say $x$ and $y$:
\begin{equation}
C(\Delta t)= \frac{<(x(t) - <x>)(y(t+\Delta t) - <y>)>}{\sqrt{<(x(t) - <x>)^2><(y(t+ \Delta t) - <y>)^2>}},
\end{equation}
where the brackets $<.>$
represent the time average; $\Delta t$ is a time shift between the dynamical variables.In our case $\Delta t=0.$ This coefficient indicates the quality of synchronization: $C=1$ means perfect complete synchronization.\\
\indent Figure 3 portrays time series of the system $z$ for parameter values $\alpha=8.01, m_{2}=m_{3}=m_{4}=m_{5}=m_{6}=m_{7}=m_{8}=m_{9}=8, m_{1}=2m_{2}=16$. Figure 4 presents synchronization error dynamics $\Delta_{z,w} =z-w$ versus time for parameters as in figure 3. $C_{z,w} =0.99$ is the cross-correlation coefficient between the systems $z$ and $w$.For parameter values as in for Figure 3 the other cross-correlation coefficients are $C_{x,y} =C_{x,z}=C_{x,u}=C_{x,w}=C_{y,z}=C_{y,u} =C_{y,w} =C_{z,u} =C_{u,w}=0.99$.\\
The value of the cross-correlation coefficients testify to the high quality chaos synchronization, which is vital for information processing in chaos-based communication systems and other possible applications.\\
It should be noted that the approach based on the Lyapunov-Krasovskii method gives a sufficient stability condition for synchronization, but does not forbid synchronization [17] when the condition (9) is not met. In Figures 5 and 6 we present the case of chaos synchronization when the stability condition for all-to-all synchronization (9) is violated. Figure 5 shows the dynamics of the system $z$ for parameter values $\alpha=3.01, m_{2}=m_{3}=m_{4}=m_{5}=m_{6}=m_{7}=m_{8}=m_{9}=8, m_{1}=2m_{2}=16$. Error $\Delta_{z,w} =z-w$ dynamics is presented in Figure 6.It is seen that despite the fact that condition (10) is violated, there is a high degree of synchronization. $C_{z,w} =1$ is the cross-correlation coefficient between the systems $x$ and $z$. For this case the other cross-correlation coefficients are $C_{x,y} =C_{x,z}=C_{x,u}=C_{x,w}=C_{y,z}=C_{y,u} =C_{y,w} =C_{z,u}=C_{u,w}=1$. \\
We notice that larger values of the relaxation coefficient $\alpha$ decrease the amplitude of the chaotic vibrations.Comparing the dynamics of the variable $z$ (Figures 3 and 5) and the error $z-w$ dynamics (Figures 4 and 6) one should pay attention to the scale on the ordinate axis.\\ 
\begin{center}
IV. CONCLUSIONS
\end{center}
\indent To summarize, we have reported on all-to-all chaos synchronization in unidirectionally nonlinearly coupled Ikeda systems. We have derived analytically the existence and stability conditions for such synchronization. Numerical simulations fully support the analytical findings. As synchronization is vital in communication systems, these results are of certain importance for information processing purposes. Additionally the results are useful for obtaining high emission power from such networks.\\
\begin{center}
ACKNOWLEDGEMENTS
\end{center}
This work was supported by the Science Development Foundation under the President of the Republic Azerbaijan-Grant No EIF-KETPL-2-2015-1(25)-56/09/1.\\

\newpage
\begin{center}
Figure captions
\end{center}
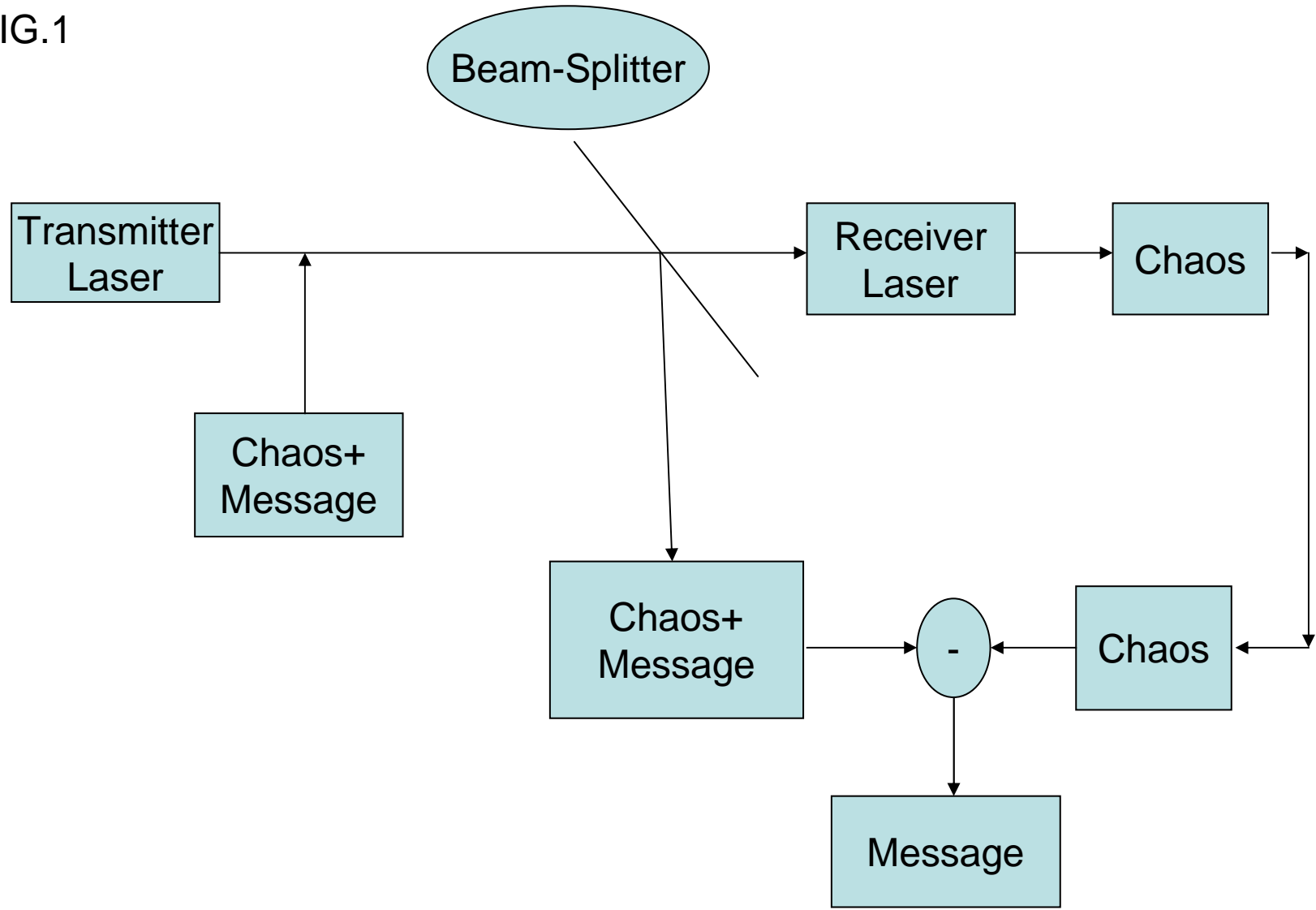
\noindent FIG.1. Schematic view of chaos based communication system. For details, see,text.\\
~\\
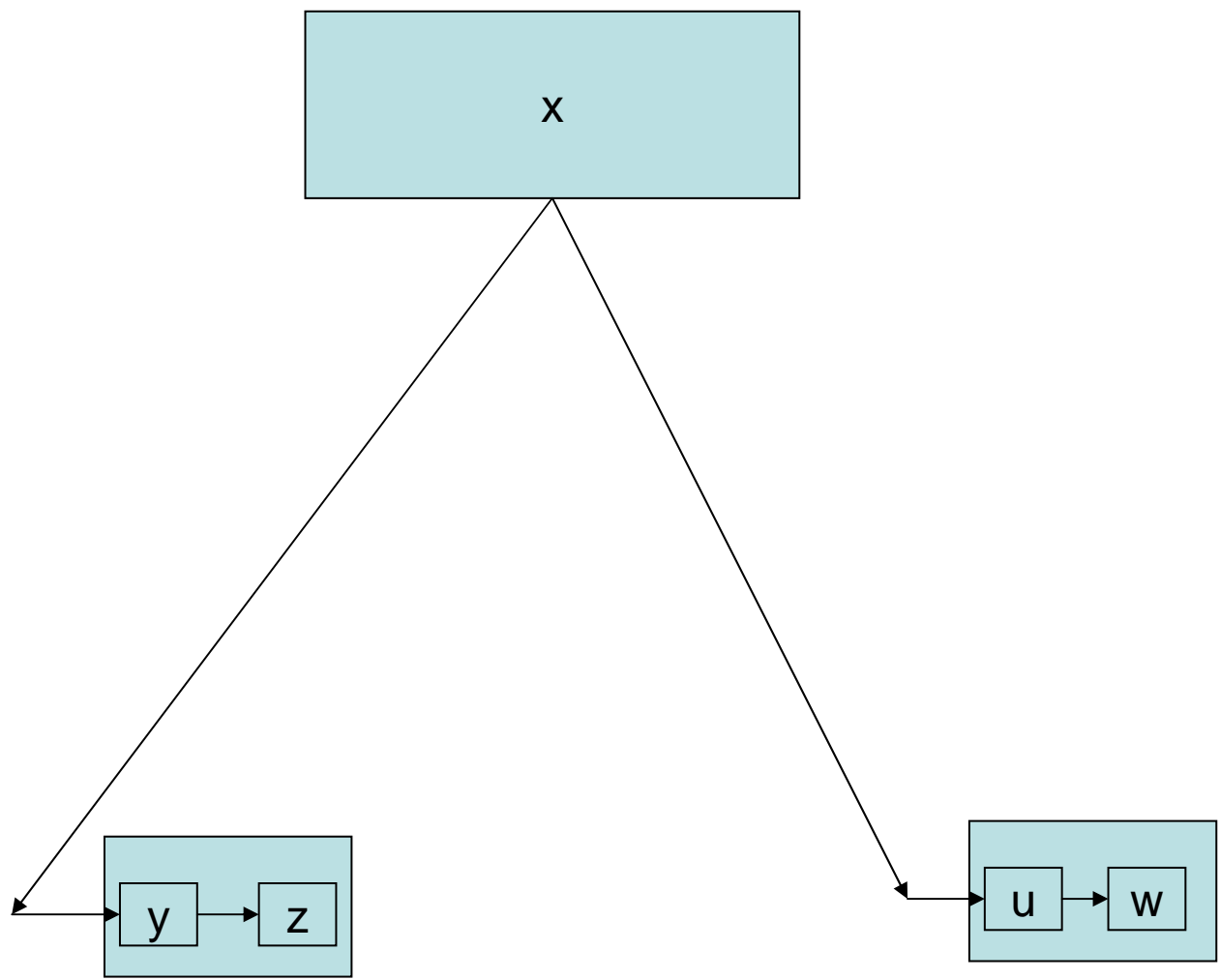
FIG.2. Schematic view of the system under consideration, see text for details.\\
~\\
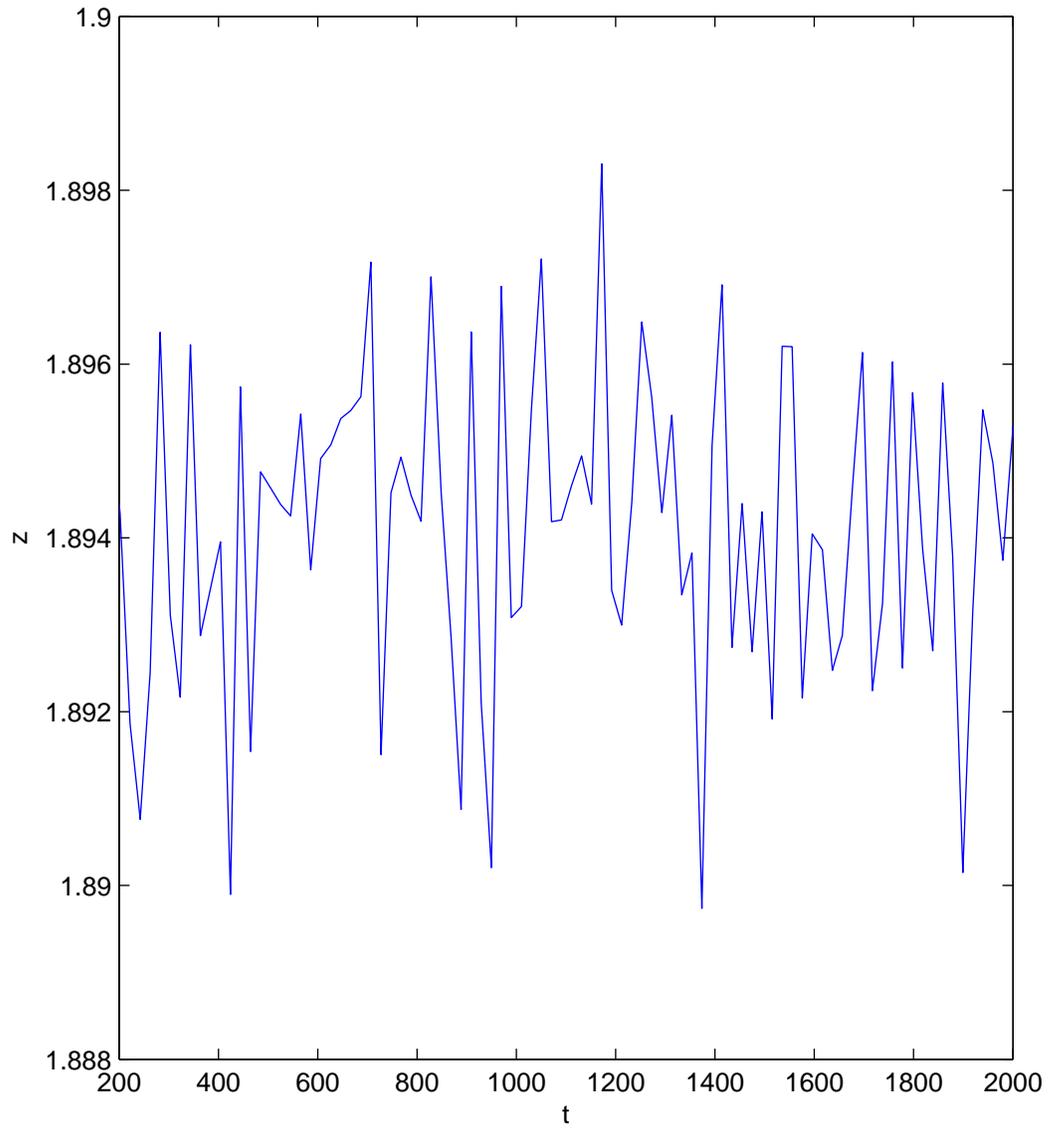
FIG.3. Numerical simulation of  all-to-all synchronization between Ikeda systems with the coupling scheme described  in Fig.2, Eqs.(1-5) for $\alpha=8.01, m_{2}=m_{3}=m_{4}=m_{5}=m_{6}=m_{7}=m_{8}=m_{9}=8, m_{1}=2m_{2}=16$.Dynamics of the system $z$ is shown. Dimensionless units.\\
~\\
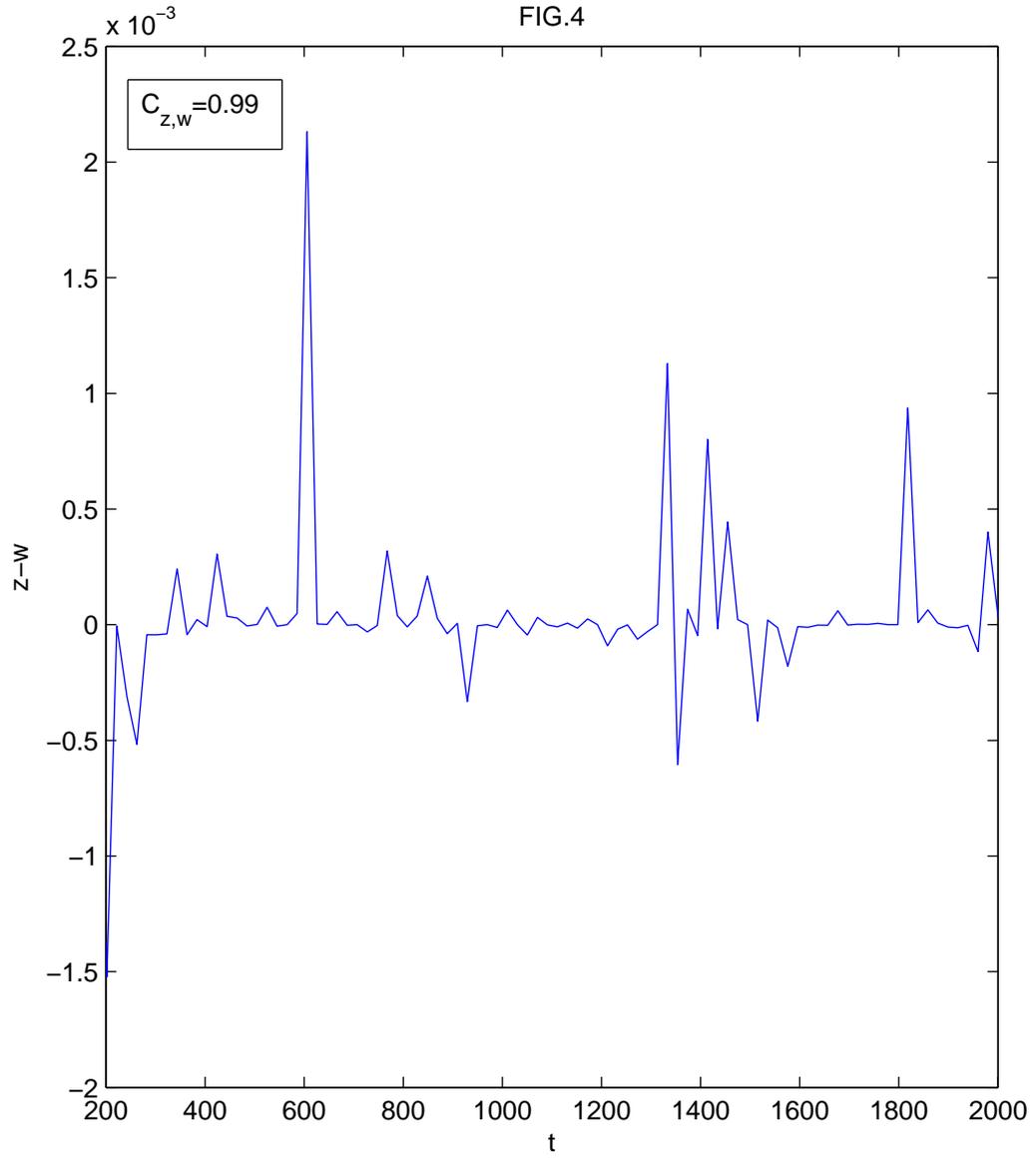
FIG.4. Error dynamics $\Delta_{z,w}=z-w$ versus time $t$ for parameters as in FIG.3. $C_{z,w}$ is the cross-correlation coefficient between the systems $z$ and $w$. Dimensionless units.\\
~\\
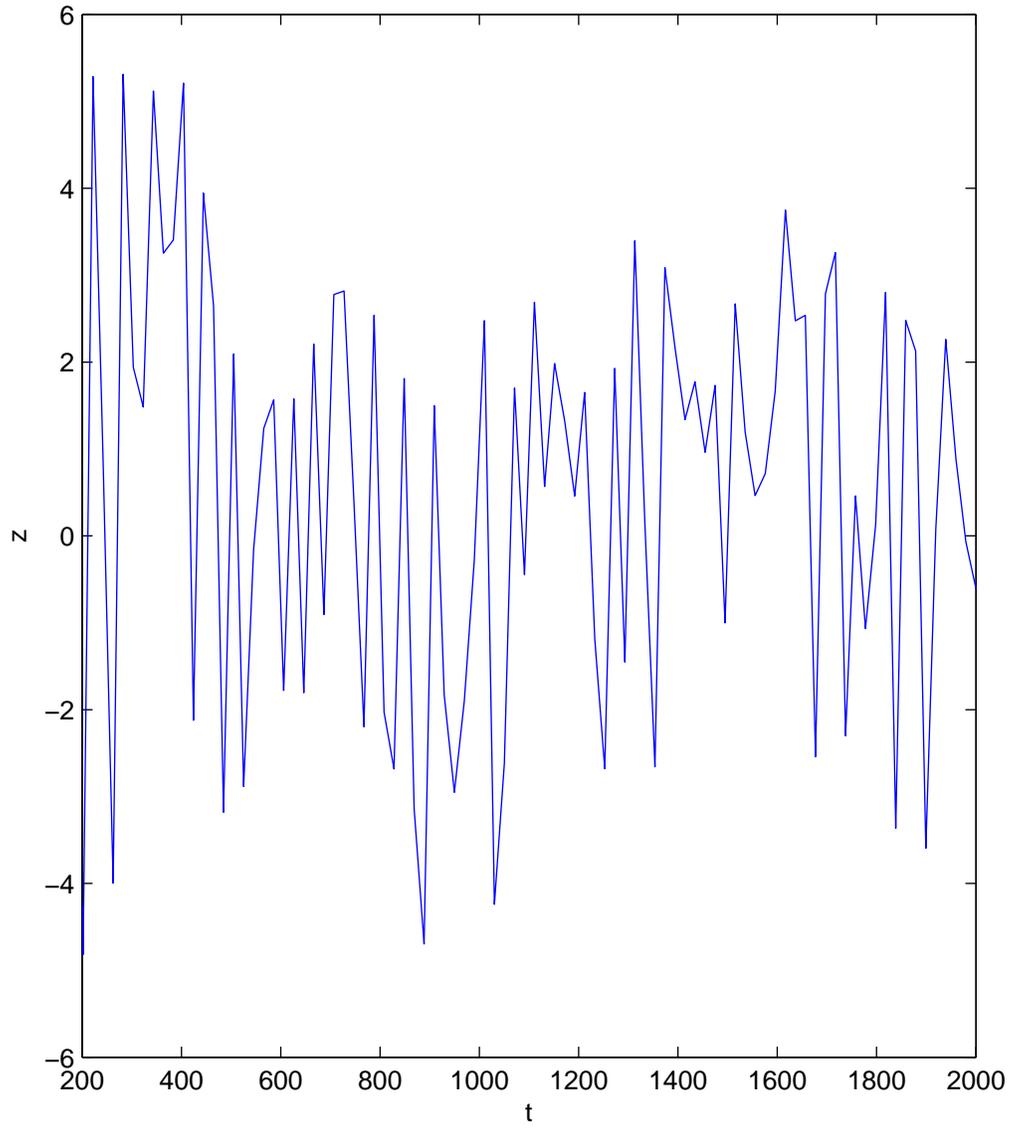
FIG.5. Numerical simulation of  all-to-all synchronization between Ikeda systems with the coupling scheme described  in Fig.2, Eqs.(1-5) for $\alpha=3.01, m_{2}=m_{3}=m_{4}=m_{5}=m_{6}=m_{7}=m_{8}=m_{9}=8, m_{1}=2m_{2}=16$. Note that stability condition (4) is not fulfilled.Time series of the system $z$ is shown. Dimensionless units.\\
~\\
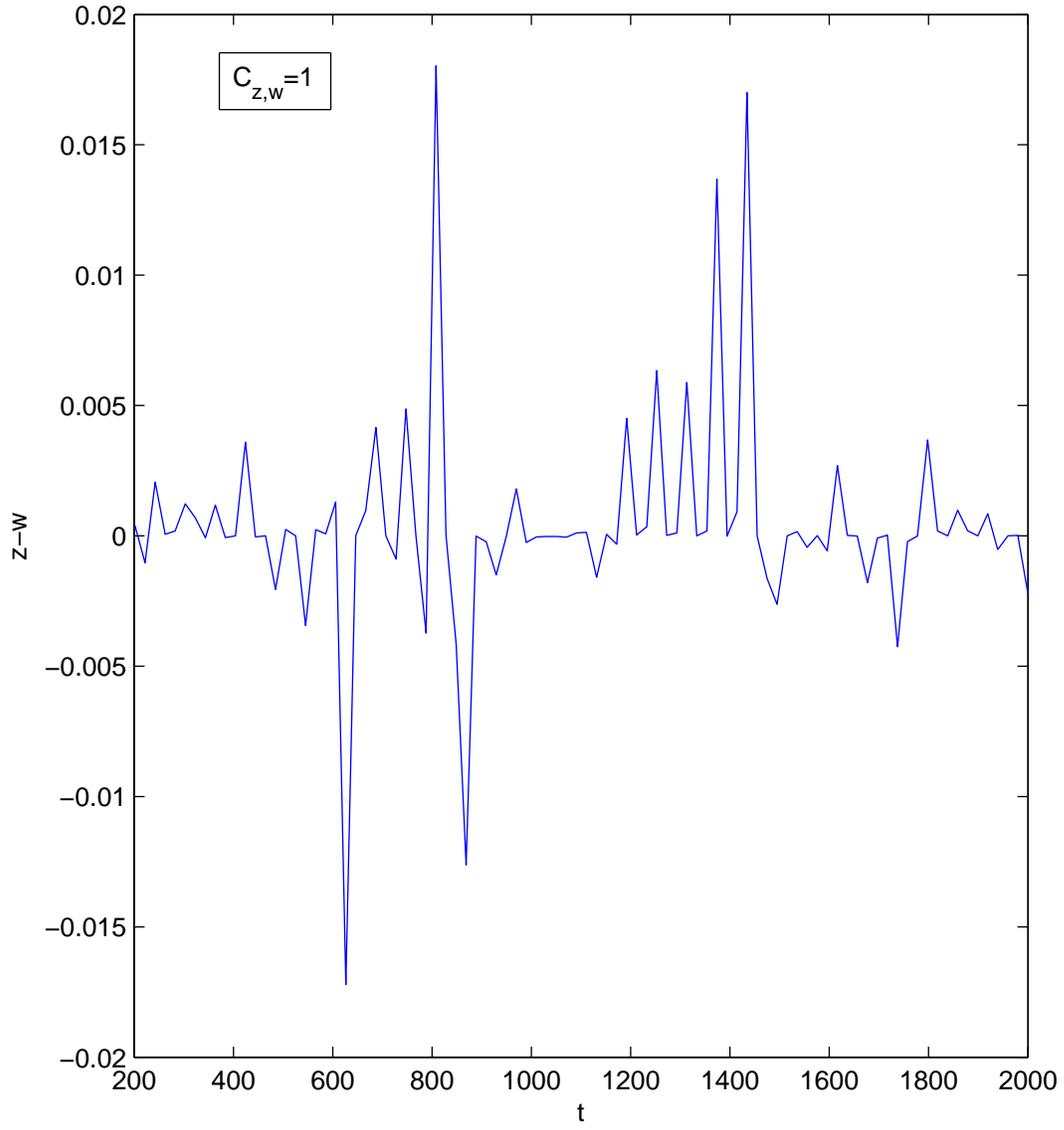
FIG.6.  Error dynamics $\Delta_{z,w}=z-w$ versus time $t$ for parameters as in FIG.5. $C_{z,w}$ is the cross-correlation coefficient between the systems $z$ and $w$. Dimensionless units.
\end{document}